# Opticle fibre calibration system and adaptive power supply[*]

**Jaroslav Cvach, Milan Janata, Michal Kovalčuk, Jiří Kvasnička, Ivo Polák and Jan Smolík**

*Institute of Physics of the AS CR,*
*Na Slovance 2, 18221 Praha 8, Czech Republic*
*E-mail:* cvach@fzu.cz

ABSTRACT: We summarize the recent activity of our group in the calibration, monitoring and gain stabilization of photodetectors, primarily silicon photomultipliers, in calorimeters using scintillator as active medium. The task originally solved for the CALICE analog hadron calorimeter founds application in other experiments.

---

**Contents**

## 1. Introduction

During the data taking with the scintillator tile calorimeter it is important to monitor the stability of the entire read-out system starting from the quality of the light transfer between scintillator and photodetector to the read-out electronics. The SiPM used as photodetectors in the AHCAL prototype [1] have gain which depends significantly on temperature and bias voltage. Therefore, the response of SiPMs to LED flashes is periodically recorded by the read-out system and this response is used off-line to correct for the gain changes. We developed several versions of the LED driver which provides light pulses several ns long and tuneable in amplitude. In section 2 we describe the latest development of the driver QMB1a and its modifications for the ECAL0 electromagnetic calorimeter in the experiment COMPASS. Further we present the fibre distribution system which brings the light flashes to the calorimeter modules. In section 3 the adaptive power supply stabilizing the SiPM gain is described in detail. The gain stability at the 1% level in the environment temperature variation between 5 and 40°C was achieved.

    In the AHCAL, each scintillator tile receives the LED light by its own optical fibre. This solution becomes cumbersome in the full scale calorimeter with several hundred thousand channels. We develop the light distribution system with notched fibres which could bring down the number of optical fibres by two orders of magnitude. The experience with the production of notched fibres is discussed in the section 4.

## 2. LED driver

The principle and properties of the LED driver developed for the beam tests of the CALICE AHCAL calorimeter are reviewed in [2]. The driver was modified for the EUDET HBU0 hadron calorimeter base unit as a one channel quasi-resonant main board driver QMB1 operated on the flyback principle of a DC-to-DC power convertor [3]. The driver has fixed pulse width (about 3.5 ns), repetition rate up to 100 kHz and variable amplitude. The pulse shape is smooth (half-sine). The driver is placed on the PCB 30 x 140 mm$^2$ large to match the size of scintillator tiles. Several boards can be combined in a system. This system is proposed as highly modular as one board behaves as a master and several others as slaves. The master board communicates with Slow-control (or DAQ) via CAN bus and the bus is also used as the distributor of the LVDS trigger signal.

**2.1 Modifications for the ECAL0 LED monitoring system of the experiment COMPASS**

In the COMPASS experiment at CERN, the new electromagnetic calorimeter ECAL0 will be installed in years 2015 – 6. The shashlik calorimeter ECAL0 [4] consists of 194 modules made as a sandwich, read by WLS fibres going through 109 lead and scintillator perforated layers.

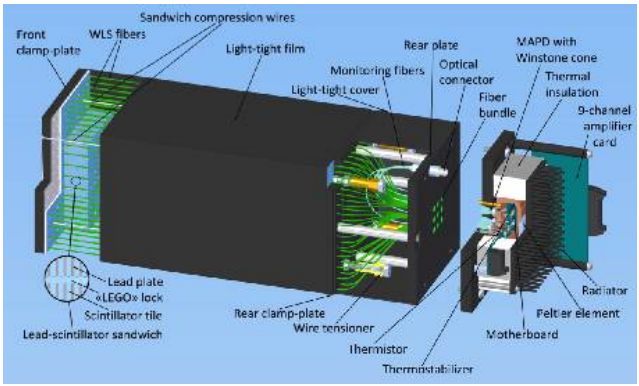

**Fig. 1.** ECAL0 module.

A layer is made from nine polystyrene scintillator tiles 40 x 40 x 1.5 mm$^3$ read by 16 fibres. One set of 16 fibres will be read by a semiconductor photodetector; both MAPD and MPPC are in consideration. Calibration light is brought by an optical fibre to each module to the optical connector and inside distributed by monitoring fibres to each photodetector (see Fig. 1).

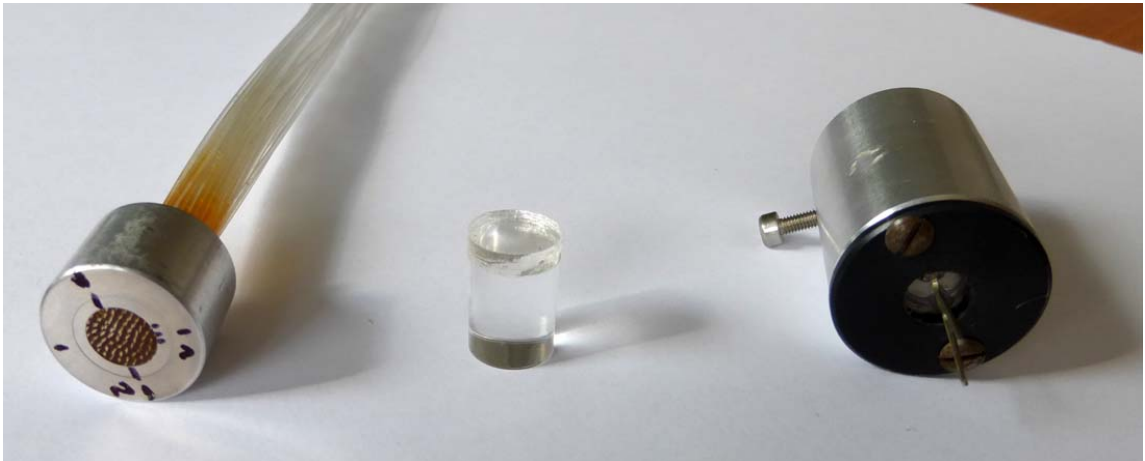

**Fig. 2.** View of the prototype of the optical bundle, light mixer and the cover.

In collaboration with the Dubna group in COMPASS (A. Nagaytsev et al.) we defined requirements on the LED monitoring system (LMS) for the ECAL0. It has two parts: LED driver and fibre distribution system. The LED driver should provide light flashes approximately 10 ns long with at least 10 000 photoelectrons in the blue light region at frequency 1 Hz. It should keep the light stable at 1% in the long term. It will be done by monitoring LED light by a PIN diode with an amplifier and averaging the light pulse intensity over approximately 10 consecutive signals. As our QMB1a uses communication via CAN bus and the experiment COMPASS uses USB or RS485 links interface CAN → USB/RS485 must be provided. The NIM signals are the standard in the COMPASS experiment whereas LVDS is used in CALICE calorimeters, therefore, also the input/output LVDS → NIM converter. The LMS should provide external and internal trigger. For the LED driver we proceed from the QMB1a with several modifications to fulfil the above mentioned additional requirements.

The light distribution system brings light from the LED to each calorimeter module. The large diameter bright LED is optically connected to a bundle of about 55 fibres one millimetre in diameter (see Fig. 2). Thus four LEDs provide light for all 194 calorimeter modules. As the LED is rather a point-like than flat source (see Fig. 3), a light mixer in the optical connector will homogenize the light to fibres.

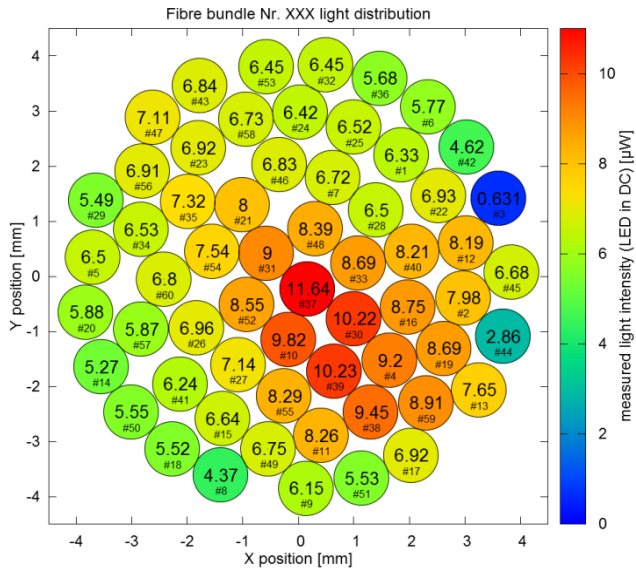

**Fig. 3.** Light distribution in a bundle of 55 fibres connected to 8 mm LED. The light intensity is given in µW. The red colour marks the spot with the highest light intensity and illustrates point like rather than flat LED source.

## 3. Adaptive power supply

The gain of SiPMs depends both on bias voltage and on temperature. To keep the gain constant, we can compensate the temperature variation by regulating the bias voltage. We have developed and built an adaptive bias voltage regulator and performed tests in a climate chamber at CERN. Over a temperature range 1 - 40 °C we have tested the performance of the bias voltage regulator with five SiPMs / MPPCs from three different manufacturers. We demonstrated that we can achieve the gain stability better than 1% for temperatures between 10-40 °C [5], see Fig. 4. In this chapter we concentrate more on the technical solution of the voltage regulator (more details given in [6]).

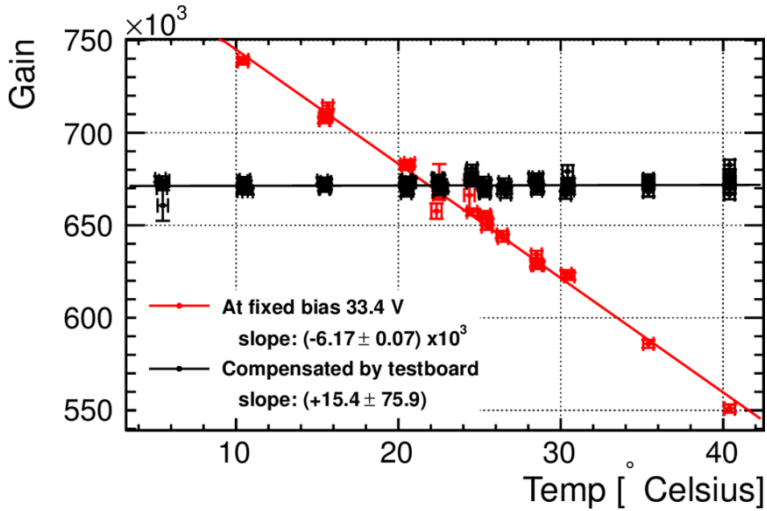

**Fig. 4.** The temperature compensated gain (black) of SiPM CPTA 857 by the adaptive power regulator. The CPTA gain without the temperature compensation is shown in red.

**3.1 The adaptive regulator**

The adaptive regulator consists of three blocks. It is HV regulator, Voltage reference and Temperature correction circuit. The supply of the regulator is low voltage 15 V and HV bulk 130V nominal. SiPM is decoupled by RC filter and connected by HV cable to regulator. Temperature sensor must be placed very close to SiPM due to affordable thermal connection with SiPM.

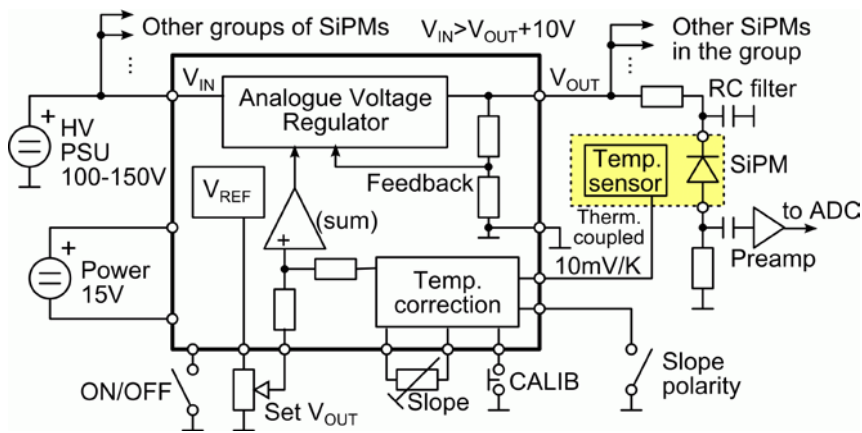

**Fig. 5.** The principal scheme of the high voltage regulator.

The power part of the HV regulator (see Fig. 5) uses two HV MOSFETs in a totem pole configuration. We implemented current limitation circuit with one NPN transistor sensing current out of the source of the upper mosfet. This limit is set to 10 mA by value of the sensing resistor. In the control part, the output voltage is divided by ten, using precision resistors[1] and attached to the feedback input of an error amplifier.

Voltage reference block is designed as a stable voltage 10.000 V what corresponds to 100.00 V at the regulator output. The LT1021 is a precision reference with ultralow drift typically 2 ppm/°C and noise 6 µV p-p (at bandwidth 0.1 Hz to 1 kHz), and extremely good long term (1000 hours) stability 15 ppms. Output voltage is set by two 10-turns potentiometers (coarse and fine tuning) in range 15 V to 100 V. We implemented a soft start circuit to get smooth ramp-up and ramp-down of the output voltage. The time constant is about 1sec. We have to wait about 10 sec. to get the output settled at $5^{th}$ digit on the voltmeter.

The temperature correction circuit is matching the signal from temperature sensor, in our case semiconductor thermometer LM35D from Texas Instruments. This sensor provides linear analogue output voltage in units of 10 mV/°C related to temperature on the smd package SO8. It is 250 mV at room temperature 25°C. This signal is then amplified 10.0 times. At the Test Point we can monitor ambient temperature on the sensor close to the SiPM, with a decimal scale 100 mV/°C and read it directly. With a 4½ digit digital multimeter we have sufficiently precise information of the SiPM temperature. For calibration purposes the circuit has the switch to ground which sets the ZERO point. To indicate +10°C, another switch connects the 100.0 µA DC current source to the 1.000 kΩ stable resistor at the input stage. We get a 100.0 mV voltage drop precisely corresponding to +10°C. The following amplifier stage further amplifies temperature signal according to the required slope. We use inverting amp to get negative polarity of the slope. We can switch between positive and negative slopes with a jumper. All current SiPMs we measured have positive compensation slope. This signal comes to the summing amplifier and is added to set the reference voltage. Output of summing amplifier is driving an Analogue HV regulator.

**3.2 Calibration and operation**

There is a simple algorithm how to set the operation point of SiPM with ADApower. We set two parameters in the system. The first one is the nominal output voltage and the second one is the slope of the correction voltage in mV/°C. We get the value of the nominal bias voltage ($V_{out}$) of SiPM from a table (photodetector datasheet). This number is normally related to room temperature 25ºC and we have to recalculate it to the value of a bias voltage at 0ºC.

We set the switch RUN to "SET ZERO" position (see Fig. 6) and then set output voltage to the value at 0ºC. Then we set other switch from ZERO to the position "ADD +10ºC". The system adds internally signal which corresponds to the increase of temperature to 10ºC (positive slope). Then we can set the slope with output voltage difference from ZERO and to +10ºC, with displayed value on the voltmeter divided by 10 (e.g., the 10ºC difference). Then we set the switch to RUN to start. Output voltage starts to swing a bit, according to the temperature drift at the sensor. It is the effect of right function of ADApower circuit, one may think of unstable malfunction at a first glance. The time constant of the feedback loop is about 100 ms. The thermal constant of a material around the sensor determines a speed of total feedback loop.

---

[1] integrated voltage divider CNS471 series by Vishay with very low tracking temperature coefficient of max.2.5ppm/ °C.

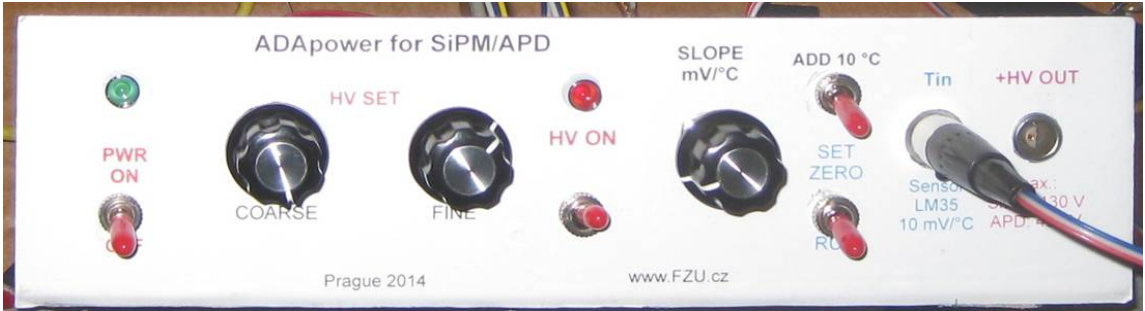

**Fig. 6.** Front panel of the control the regulator

### 3.3 Measurements

We monitor the output voltage with the 6½ digits multimeter (HP3458A, Keithley 2001, or HP34401). To check the stability in short time we used a freeze spray can (-40 ºC) to cool and 100 W lamp to heat the board to about 50ºC. Even if this is not a scientific approach due to the short time before another laboratory test, we found a scale of a drift on the output voltage in the range of 60 mV. This corresponds to the requirements for the stability of the circuits better than 10 ppm/ºC. A way to further improve the stability of the adaptive regulator is still possible, but not so easy. The key factor is to shield the voltage reference from ambient air turbulence. Air movement can create low frequency noise (< 1 Hz) because of thermoelectric differences between IC package leads and printed circuit board materials. The same is valid even more for the temperature coupling between SiPM and temperature sensor. Second factor is to use more stable resistors (better than 50 ppm/K) in signal trace and namely to use an output voltage divider with better long term stability. The measurement of the gain stabilization of SiPM proves the functionality and parameters of developed circuit (see Fig. 4).

## 4. Notched fibre light distribution system

The high granularity and compactness of the AHCAL calorimeter are demanding on the size and simplicity of the light distribution system. The solution used in case of the first AHCAL prototype, where one LED flash through bundle of 16 fibres into 16 tiles is not applicable for the future calorimeter. There are two solutions – direct application of LED driver into PCB lying over tile [7] or using external LED driver with optical fibre for several tiles. Each of this solution has advantages and disadvantages which should be taken into account. The first one by avoiding installation of fibres significantly simplifies manufacturing of calorimeter. On other side the number of drivers to be tuned is very high, it is equal to the number of tiles, and flashing of all the LED drivers closed in steel absorber construction can cause significant cross-talk with readout electronics located on the same PCB. The second solution needs manufacturing and installation of fibre distribution system but on other side simplifies tuning and operation of the system.

To distribute the calibration light into several tiles the idea of "notched fibre" is used. It is a standard POF fibre of 1mm diameter (can be changed depending on the application) which is controlled damaged on defined points. In the notched point the part of light is dissipated around the fibre. The tests show that the light is dissipated with maximum on the opposite side to damaged spot. To test the idea, three fibres with 72 notched points at 3 cm distances were hand-made. The notches were produced by round file. The results were satisfactory and it was shown that this system can work. The signal generated in scintillating tile by light from most of individual notched points did not differ by more than 30% from average value. At maximum five points on each fibre were out of this limit.

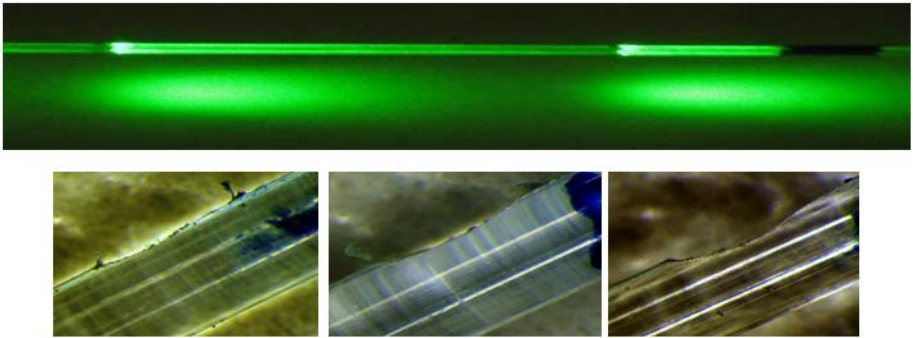

**Fig. 7.** View of the light spots irradiated from notches in the fibre (upper). A detailed view of the notch at the beginning of the fibre (close to the LED), in the middle and at the end of the notched fibre (lower). To keep the light intensity constant along the fibre, the notch size increases toward the fibre end.

Next, the development of semi-automatic procedure was started. The production was tested using simple 3-dimensional CNC machine ML 1000F from CNC-multitool. Produced fibres had only 24 notches, to cover the whole length of the calorimeter module with 72 tiles three fibres are used. This solution allows emit from one notched point more light than in case of a 72 notches fibre. Fibre is fixed in a special template and notched points are milled by milling cutter of 3 mm diameter. The template allows to measure the emitted light during processing in similar condition as in the experiment. The milling depth is gradually increased till the given level of emitted light is reached. After that the next point is milled down stream from the light entry to the fibre end.

In all, several tens of fibres were produced with several modification of milling algorithm. The spread of ±20% around average value was achieved but still one or two points are significantly over of this limit. Usually these points are at the beginning of the fibre where the notched points are shallowest. It is due to the coarseness of the CNC machine minimal step which is 10 μm. To achieve spread of ±10% in emitted light we estimate the needed minimal machine step to 1-3 μm. The recent production time for one fibre is about 1-2 hours.

Regardless of it, the full calibration chain – QMB1a LED driver with bundle of three 24 notches fibres were successfully tested at DESY. The whole 72 tiles row (except 12 tiles which were absent) was calibrated by the system. The single photoelectron peaks spectra and saturation curve of SiPM were measured.

Based on the current experience with semi-automatic procedure the technical proposal for dedicated production machine is prepared. It is based on the milling head which will allow to fix fibre, measure the emitted light, mill in steps at the level of 1 μm and with milling cutter of diameter less than 3 mm and to measure the distance between milling cutter and the fibre. The last ability can improve the production rate – estimation is 30 s for one point production. The fibre will be carried through the milling head. Unlike to the template for fixing fibre it will allow to produce fibre with arbitrary distance between notched points.

## 5. Conclusions

We reviewed the recent activities of our group in calibration and monitoring with LEDs for scintillation calorimeters using SiPMs as photodetectors. The improved QMB1a LED driver with the PIN feedback is now developed for the LED monitoring system for the electromagnetic calorimeter in the COMPASS experiment. The experience with the production of notched fibres led to the survey of the whole process with the milling of notches. We have a proposal on the design of a new milling machine to reach the goal – the spread of light intensity radiated from 24 notches less than 15%. We finished the development of the adaptive power supply which stabilizes the gain of SiPMs at the 1% level in the temperature range 5 – 40°C. It could be advantage to control the Adaptive power supply (ADApower) by a computer. We can modify the circuit and add an ADC/DAC control ability. The next task for the temperature compensation of the SiPM gain is to extend the power supply to a calorimeter with hundreds of SiPMs.

## Acknowledgments

We gratefully thank to our colleagues from JINR Dubna A. Nagaytsev et al. for the support during the development of LMS for the ECAL0 of the experiment COMPASS, the SAFIBRA firm in Ricany u Prahy, Czech Republic for the technical assistance during the production of notched fibres and to our colleagues from the University of Bergen led by G. Eigen for the measurement of temperature gain dependence of SiPMs and the support during the adaptive power supply development and production. This work was supported by the Ministry of Education, Youth and Sports of the Czech Republic as project LG14033 and by the EC as INFRA project no. 262025, AIDA.